# Can Artificial Intelligence solve the blockchain oracle problem? Unpacking the Challenges and Possibilities

Giulio Caldarelli

*University of Turin*

giulio.caldarelli@unito.it

ORCID ID: 0000-0002-8922-7871

**Abstract**

The blockchain oracle problem, which refers to the challenge of injecting reliable external data into decentralized systems, remains a fundamental limitation to the development of trustless applications. While recent years have seen a proliferation of architectural, cryptographic, and economic strategies to mitigate this issue, no one has yet fully resolved the fundamental question of how a blockchain can gain knowledge about the off-chain world. In this position paper, we critically assess the role artificial intelligence (AI) can play in tackling the oracle problem. Drawing from both academic literature and practitioner implementations, we examine how AI techniques such as anomaly detection, language-based fact extraction, dynamic reputation modeling, and adversarial resistance can enhance oracle systems. We observe that while AI introduces powerful tools for improving data quality, source selection, and system resilience, it cannot eliminate the reliance on unverifiable off-chain inputs. Therefore, this study supports the idea that AI should be understood as a complementary layer of inference and filtering within a broader oracle design, not a substitute for trust assumptions.

1. Introduction

"*As Bitcoin nodes cannot measure arbitrary conditions, we must rely on an 'oracle*" (Mike Hearn 2011[1]). Blockchain technology promises decentralized, secure, and transparent interactions, thereby reducing or eliminating reliance on trusted third parties [2], [3]. However, beneath this appealing aspect lies a crucial, unsolved issue: the so-called "blockchain oracle problem [4], [5]." At its core, the oracle problem reflects a fundamental challenge of blockchains that, although powerful for ensuring computational trust and consensus on internal state, are inherently incapable of verifying the correctness of external information fed from the real world [6]. Thus, blockchains must rely on external entities, so-called "*oracles*" to bridge on-chain computation with off-chain reality [7]. These oracles, in turn, reintroduce an unwanted layer of trust into systems intended to be trustless, effectively becoming single points of failure and manipulation [8].

In recent years, artificial intelligence (AI) has rapidly gained traction as a disruptive technology, celebrated for its ability to analyze vast datasets, detect anomalies, predict events, and even perform automated reasoning tasks with increasing accuracy [9], [10], [11]. Given AI's rising significance across industries, it is natural to consider whether this powerful tool could finally address or even solve the persistent blockchain oracle problem. This paper aims to clarify that position by moving beyond the current hype and providing a balanced analysis of AI's strengths and shortcomings in oracle infrastructures. By doing so, it intends to inform future research and promote more robust oracle architectures. Drawing from technical literature and practical implementations, we analyze how AI methods, ranging from anomaly detection and reinforcement learning to large language models, can be applied to oracle design. We investigate the realistic potential of AI to support blockchain oracle systems, critically exploring whether AI can mitigate or even solve core vulnerabilities such as data reliability, source trustworthiness, and systemic manipulation.

The study reveals that recent developments in oracle infrastructure have begun incorporating AI techniques at multiple levels. For instance, protocols such as Supra have proposed Threshold AI systems where oracle nodes are powered by AI agents that lock collateral and are rewarded or penalized based on performance, using reinforcement learning principles to enhance data reliability and responsiveness [12]. Meanwhile, Chainlink has explored AI-driven risk scoring for oracle reputation, while Oraichain integrates AI APIs into smart contracts for on-chain inference, combining machine learning with blockchain-native validation processes [13], [14]. In parallel, academic proposals such as AiRacleX leverage large language models to proactively detect oracle manipulation attempts in decentralized finance protocols by extracting known vulnerability patterns from blockchain literature and analyzing smart contracts accordingly [15]. Similarly, research frameworks like SenteTruth aim to standardize the use of LLMs in oracle contexts by enforcing deterministic output behavior across nodes and verifying consistency through multi-model consensus strategies [16].

These integrations suggest that AI can increase oracle accuracy, adaptability, and efficiency, but, as further discussed in Sections 3 and 4, they also introduce new risks related to non-determinism [16], [17], hallucination [18], bias [19], adversarial manipulation [20], [21], and architectural complexity [22]. We therefore argue that AI, while valuable, cannot fully solve the oracle problem, as the issue is not just technical but epistemological. AI models, regardless of sophistication, rely on the integrity of their inputs, making them susceptible to the same trust limitations oracles face. Therefore, the use of AI in oracles should be framed not as a solution, but as a complementary layer within a broader system of cryptoeconomic guarantees, governance rules, and verifiability mechanisms.

The paper proceeds as follows. Section 2 introduces the blockchain oracle problem and reviews the evolution of oracles and artificial intelligence. Section 3 explores AI techniques and how they can be integrated into oracle design. Section 4 presents the main limitations of AI when applied to oracle systems, emphasizing the persistence of the underlying problem. Section 5 concludes with final remarks and research directions.

## 2. Literature Review

This section provides a comprehensive overview of blockchain oracles and the oracle problem, tracing the historical development and technical evolution. It further reviews foundational concepts in artificial intelligence, including expert systems, machine learning, reinforcement learning, NLP, and adversarial robustness, to establish the necessary background for understanding and evaluating AI-based oracle solutions proposed in later sections.

### 2.1. Defining the Blockchain Oracle Problem

The blockchain oracle problem emerges directly from blockchain technology's inherent limitation, as the inability of blockchain systems to independently verify external, real-world data [23]. While blockchain's core innovations, such as immutability, transparency, and decentralization, make it ideally suited to create trusted, cryptographically secure environments, the utility of blockchains in real-world applications critically depends on external data integration [4]. As blockchain networks cannot inherently confirm the authenticity of this extrinsic data, they must rely on intermediaries known as oracles to bridge this gap. Arguably, while an ideal bug-free smart contract may allow the exchange of cryptocurrency trustlessly, the exchange rate of these currencies, which is a piece of data pertaining to the real world, needs to be provided by an oracle whose credibility and reliability cannot be ensured [4], [5], [8]. The oracle problem thus refers to the fundamental contradiction between the need for trusted external inputs to feed blockchain systems, which inevitably reintroduce elements of centralization and trust, and blockchain's foundational goal of decentralization and trustlessness [6], [7]. Previous studies have also distinguished various dimensions of the oracle problem, as failure to provide reliable data may be either due to technical difficulties, in case of bugs, tampering, or poor programming in good faith. On the other hand, social matters may also affect the reliability of oracle data due to competing interests of oracle managers or other malevolent actors [24], [25], [26], [27].

### 2.2. Historical Context and Evolution of Oracles

Initial attempts to integrate real-world data into blockchains began with early Bitcoin experiments, where developers confronted immediate and profound limitations due to Bitcoin's rigorous decentralization principles. Interviews with early Bitcoin developers reveal that, already in the early days, the concept of introducing external data into blockchain systems was met with skepticism and described provocatively as "cheating," highlighting early recognition of the tension between complete decentralization and practical functionality [28]. The same Nakamoto was skeptical on the idea of integrating oracles and advocated for alternative solutions, which, however, didn't manage to implement by the time he left the scenes [29].

With the advent of Ethereum EVM, however, and the massive financing of decentralized apps, the delays in oracle problem solution would have negatively impacted the wave of decentralized innovation brought by web3 and the relative financing; therefore, blockchain integrations were made with the tentative oracles available at that time [30]. This hyped, driven rush to oracles implementation aroused mixed feelings from those who advocated for further testing and a more cautious approach [23], [28], [31]. Arguably, an excess of prudence may have negatively impacted blockchain innovation; however, the billions of dollars lost in DeFi due to oracle manipulation are a clear sign that a more cautious approach could have been adopted [32], [33], [34].

Despite major advancements in blockchain technology, the fundamental conceptual problems of oracles identified in the early Bitcoin era persist, and their development has remained a critical yet often overlooked challenge within blockchain literature and practice [38].

When it comes to blockchain oracles solutions, they differ greatly in structure, reliability, and purpose. Various classifications are offered in literature that evolve in parallel with new solutions developed [24], [27]. The most basic form is constituted by a centralized oracle: a solution proposed in the first days of Bitcoin with the aim of just enabling real-world integrations. In this phase, we still don't have official oracle protocols, as they were mostly prototypes developed ad hoc, based on legacy computer science. As they reintroduced single point of failure and other issues as unverifiable data, an early oracle protocol named Oraclize, leveraged Trusted Execution Environment and cryptographic proof to guarantee that the data provided came from a trusted source and was not manipulated [39]. However, this did not address the oracle problem completely, as it still reintroduced a single point of failure. Although a cryptographic proof could indeed prove that the data was not altered in the delivery, it cannot prove that the data was truthful. For that reason, alternatives were proposed, such as Orisi (still on Bitcoin), whose intent was to enhance decentralization by implementing multiple data reporters to ensure that no single actor could manipulate the data requested [40]. Additionally, although this solution was strong in design, it still couldn't solve the problem. Orisi voters could in fact collude due to competing interests and being anonymous they could be easily replaced by the same agent impersonating multiple entities. This condition extensively discussed and described in literature is known as Sybil attack [41]. The initial approach to address this problem was constituted by decentralized oracles based on game-theoretical models such as Truthcoin [42]. The rationale was to make it inconvenient for agents to deceive the system and always provide honest information. Although groundbreaking, the limit of these systems is, however, the theoretical and implementation complexity; it is enough to say that Truthcoin, conceptualized in 2014, is still in development nowadays.

When it comes to technical integrations, the first oracles, developed on Bitcoin, primarily used multi-signature techniques and conditional scripts to introduce external data, employing manual or semi-automated processes. These approaches evolved significantly with Ethereum, enabling automated, sophisticated smart contracts capable of consuming complex external data streams via APIs and introducing tokens as incentives. Early days protocols, such as Augur and Witnet, involved, in fact, the use of tokens as a representation of a reporter's reputation. However, although representing an intriguing idea, the token management still brings some challenges. Reputations stacked with tokens can easily be sold, stolen as in the case of Augur, or lost for inactivity in environments such as Witnet. Ethereum however allowed also another interesting solution known as First-Party oracle [43], [44]. The idea, developed by API3, allows any entity to become a blockchain data provider through software that in the case of API3 is named "airnode". That way, trusted entities of the real-world can provide their data, dramatically enhancing the reliability of web3 implementations. Although disruptive, this idea also has limitations as it can't remove the risk of failure and data manipulation. Finally although facilitating the process it cant oblige any entity to be a Web3 data provider [45]. The advent of alternative blockchains further highlighted the issue of interoperability. Since blockchains are inherently isolated systems, unable to natively access external data, they are also unable to communicate directly with one another, thereby extending the oracle problem to inter-chain communication [35], [36], [37].

Table 1 provides an overview of available oracles and related advancements/drawbacks

Table 1. Blockchain Oracles' evolution with pros and cons.

| Type / Architecture | Key Mechanism / Technology Used | Primary Use Case / Domain | Oracle Problem Dimension Addressed | Notable Implementations / Examples | Limitations or Open Issues | Reference(s) |
|---|---|---|---|---|---|---|
| Centralized Oracles | Single trusted data source | Simple dApps, early DeFi | Latency, cost efficiency | Early Bitcoin integrations | Single point of failure, unverified data | [4], [26], [30] |
| Multi-source Aggregator Oracles | Aggregation of multiple sources with voting | DeFi, stablecoins, derivatives | Minimizing manipulation, improving consensus | Orisi, Band Protocol, Nest | Still vulnerable to collusion, sybil attacks. | [40], [46], [47] |
| Reputation-based Oracles | Reputation scoring for source selection | Insurance, games, forecast markets | Trust calibration, fault tolerance | Augur, Witnet | Hard to quantify or update reputation objectively | [43], [44] |
| Crypto-economic Incentivized Oracles | Token-based staking, slashing, or dispute systems | Synthetic assets, prediction markets | Incentive alignment, manipulation resistance | Truthcoin, UMA, Tellor | Complexity, risk of game-theoretic exploits | [42], [48], [49] |
| TEE-based Oracles | Trusted Execution Environments (Intel SGX) | High-value finance, sensitive data | Data integrity, confidentiality | Oraclize, Town Crier, DECO | Hardware trust assumptions, opaque execution | [39], [50], [51] |
| First-party Oracles | Direct data from original source owner | Enterprise oracles, IoT, insurance | Data origin authenticity | API3, Chainlink OCR | Scaling, need for robust standardization | [45], [52] |

*Author elaboration

## 2.3. Why the Oracle Problem Persists

Despite technological advancements, the fundamental oracle problem persists primarily because it is inherently epistemological rather than purely technical. Blockchain's inability to independently verify external truths implies that oracles as intermediaries inherently reintroduce trust dependencies into systems initially designed to eliminate them. This creates a paradoxical dynamic in blockchain applications since while blockchain can guarantee the integrity of data once on-chain, it cannot guarantee the veracity of external data introduced onto the chain.

Moreover, blockchain literature frequently underestimates the depth of this problem, often assuming ideal conditions for oracle functionality. In fact, numerous articles and research demonstrate the lack of global interest and contribution to the subject and the dramatic negative consequences that this oversight is causing [4], [5], [6], [7], [38], [53]. Major blockchain integration projects, such as the case of Maersk, are progressively being abandoned since, if a higher degree of decentralization cannot be achieved due to the reliance on third parties, the integration of blockchain would inevitably result in a slow and overcostly legacy application [54], [55].

The significance of the blockchain oracle problem is substantial across numerous industries. Applications ranging from decentralized finance (DeFi) and supply chain tracking to insurance and governance critically depend on oracle integrity. However, the literature emphasizes that the severity and practical impact of the oracle problem are context-specific: the more critical the system's reliance on unverifiable external data, the more profound and risky the oracle problem becomes [56]. On the other hand, while the oracle problem as a whole cannot be addressed, specific solutions may prove to be particularly effective in some applications and guarantee a satisfactory level of reliability.

TWAP, for example, is a technique that can efficiently address outliers in applications such as price feeds [57]. In DeFi, for example, oracles are asked to provide asset prices in order to perform exchanges or evaluate the value of an investment for lending or borrowing purposes. If an asset is instantly mispriced as overvalued or undervalued, a malevolent agent can drain value from a liquidity pool, exploiting the price difference, or users can see their position liquidated as the protocol can detect their lending position as undercovered. Cases such as the Compound incident or the dydx hack are examples of these situations [58], [59], [60]. The TWAP technique helps prevent these unwanted events by performing a time-weighted average of price feeds. Instead of digesting extrinsic data directly from the source, the oracle performs an average that allows for abnormal feeds to be dropped, thus avoiding the above-mentioned consequences. Interestingly enough, the TWAP does not solve the oracle problem as unwanted events are prevented by dropping outliers, but outliers are not necessarily wrong values. Plus, the mean value is not necessarily the true value of an asset in that place and time [61].

Another case is represented by the use of a simple digital signature that can dramatically improve blockchain-based notarization or academic records. Arguably, due to the oracle problem, the blockchain cannot verify the authenticity of a document, nor can it ensure that the person uploading has the legal right to upload it to the chain. However, the use of a digital signature may efficiently address both these issues. This technology unequivocally links a digital file to a person or company, therefore ensuring its ownership. Academic records uploaded on the Bitcoin network by the MIT University, for example, will have the MIT digital signature and therefore will be recognized as authentic [62], [63]. Again, blockchain cannot prevent any other entity from uploading a false MIT certificate on the chain. However, this entity cannot sign the certificate with the MIT digital signature and therefore will be identified as nonauthentic, regardless of its quality or content. Furthermore, even if a certificate is uploaded by MIT University itself, the fact that is on the chain does not ensure that the content is true. However, the hypothesis that a renowned institution uploads a fake certificate is evidently hilarious. On the other hand, if an institution that is not trusted uploads documents on the blockchain, this will not affect its trustworthiness; therefore, the oracle problem is not the real issue in this case [56].

The above-mentioned examples support the view that although not solving the oracle problem, there is evidence of robust solutions that can allow blockchain integrations to be reliable up to a certain extent.

Since the recent improvement in artificial intelligence is positively impacting numerous sectors it is plausible to hypothesize that certain AI applications can positively address aspects of the oracle problem resulting in more reliable blockchain integrations. But, in order to correctly speculate AI integration in oracles we believe it necessary to provide a thorough literature background on AI applications as current overhyped narrative on this technology (as it was for blockchain during its bubble), may alter its correct positioning. Therefore, with the following section, we aim to give an idea of the AI, its real potential in solving conundrums, and, where possible, implications will already be introduced for possible oracle integrations.

### 2.4 Early Symbolic AI and Expert Systems

Modern artificial intelligence traces its roots to mid-20th-century visionaries. In 1950, Alan Turing famously posed the question *"Can machines think?"* and introduced the Imitation Game (later known as the Turing Test) as a benchmark for machine intelligence [64]. Just a few years later, the field was formally born at the 1956 Dartmouth workshop organized by John McCarthy who conjectured that *"every aspect of learning or any other feature of intelligence can in principle be so precisely described that a machine can be made to simulate it."*[65]. This bold assumption is at the base of early symbolic AI research, which sought to encode human knowledge and reasoning in machines using formal logic and symbols.

By the 1960s and 1970s, symbolic AI had yielded expert systems programs that emulated the decision-making of human specialists through explicit rules. A landmark example was MYCIN, a rule-based medical diagnosis system at Stanford. MYCIN leveraged more than 600 handcrafted rules to identify blood infections and recommend related medications such as specific antibiotics [66]. Impressively, in blinded evaluations, MYCIN's therapeutic recommendations were slightly preferred by medical judges over those of human infectious disease experts [67]. This demonstrated that codifying expert knowledge could achieve expert-level performance in narrow domains. Other notable expert systems (e.g., DENDRAL for chemistry, PROSPECTOR for geology) likewise showed high accuracy within specialized tasks [68], [69]. However, these systems lacked generality and common-sense knowledge, a limitation famously pointed out by McCarthy and others who argued that broader *"common sense"* reasoning was needed beyond narrow rules [70]. The knowledge engineering effort to manually curate facts and rules proved difficult to scale, setting the stage for new approaches.

**2.5 Machine Learning and Neural Networks**

In the 1980s and 1990s, AI shifted toward machine learning (ML) algorithms [71] that enable computers to learn patterns from data rather than rely on fully hand-crafted rules. An early precursor was the Perceptron [72], a simple neuron-like model that learned to classify inputs. While a 1969 analysis by Minsky and Papert highlighted perceptrons' limitations (triggering a temporary retreat from neural approaches), the subsequent decades saw a relaunch of "connectionist" ideas [73]. A pivotal breakthrough came with the re-discovery of the backpropagation training algorithm by Rumelhart et al. [74]. Backpropagation allowed multi-layer neural networks to adjust their weights to minimize errors, enabling these deep neural networks to automatically learn useful internal representations of data. Basically, the error was calculated between the network prediction and the actual result. The information is then distributed "backward" in the network with the aim of reducing future errors. This overcame earlier training obstacles and unlocked greater modeling power than single-layer perceptrons. Rumelhart et al.'s work along with parallel advances in computing reignited neural network research and led to rapid progress in pattern recognition tasks.

In parallel, statisticians and computer scientists developed powerful non-neural ML methods. Decision trees, Bayesian networks, and support vector machines [75], were also developed as methods that learn how to make decisions or separate different types of data by looking at examples. By the early 2000s, countless ML algorithms were already available, and data-driven learning had firmly established itself as part of the AI spectrum. Over the next decade, neural networks scaled up in depth and data, initiating the deep learning revolution. Hinton and collaborators demonstrated that deep neural networks could be pre-trained layer-by-layer (e.g. via restricted Boltzmann machines, [76]), overcoming optimization difficulties. In 2012, a deep convolutional network by Krizhevsky, Sutskever & Hinton [77] dramatically improved image recognition benchmarks, halving the error rate on the ImageNet challenge [78]. Such results catalyzed widespread adoption of deep learning across vision, speech, and beyond. By 2015, deep learning achieved state-of-the-art performance in numerous domains, from image classification and object detection to speech recognition and drug discovery. Goals of deep learning include multi-layer feature learning and end-to-end optimization via backpropagation, yielding systems that often outperform earlier symbolic or linear models by discovering specific structures in large datasets.

## 2.6 Reinforcement Learning

Another pillar of AI is reinforcement learning (RL), a principle inspired by behavioral psychology, where an agent learns to make decisions through trial-and-error interactions with an environment. In an RL framework, an agent receives rewards for desirable outcomes and seeks to maximize its cumulative reward by improving its policy (behavior strategy) over time [79]. Sutton and Barto's seminal work formalized RL methods and algorithms, establishing it as a distinct research area in the 1980s and 1990s. Key advances included temporal-difference learning [80] and Q-learning [81], which enabled agents to learn value functions and optimal policies from delayed feedback.

Classic achievements in RL involved game-playing and control: for example, Tesauro's TD-Gammon [82] learned to play backgammon at world-champion level through self-play, and in the 2010s deep RL methods produced striking results like DeepMind's AlphaGo [83] defeating human Go masters. These milestones illustrate the power of combining reinforcement learning with deep neural networks (deep RL) for complex sequential decision problems. In the context of blockchain oracles, RL's contribution may be more indirect (e.g. training autonomous agents to optimize data sourcing strategies or adaptively choose information sources). Nonetheless, RL's core idea of learning behavior by reward feedback could inform oracle mechanisms that self-tune based on successful outcomes (e.g. rewarding oracle nodes for accuracy or usefulness of provided data).

## 2.7 Natural Language Processing and Large Language Models

Natural Language Processing (NLP) has been for long a challenging AI domain, requiring an understanding of human language's complexity and ambiguity. Early NLP systems in the 1960s–1980s (like ELIZA or SHRDLU) used rule-based or grammar-based techniques, but they were naive and domain-limited [84], [85]. The 1990s brought a shift to statistical NLP, leveraging probabilistic models and corpora (for instance, n-gram language models or Hidden Markov Models for speech). This era saw improvements in machine translation and speech recognition by treating language tasks as ones of pattern recognition on large text datasets.

The true revolution in NLP came with the advent of neural network approaches. Recurrent neural networks (RNNs) and their variants (e.g. LSTM networks by Hochreiter & Schmidhuber [86]) allowed modeling of sequential language data with memory of prior context, greatly improving tasks like handwriting recognition and translation. By 2014, sequence-to-sequence (Seq2Seq) models with RNNs [87] and attention mechanisms [88] enabled end-to-end neural machine translation, surpassing traditional statistical methods. This culminated in the breakthrough Transformer architecture proposed by Vaswani et al. [89], which replaced recurrence entirely with multi-head self-attention mechanisms. The Transformer proved more efficient to train (highly parallelizable) and achieved superior accuracy on translation benchmarks, establishing a new state of the art. It has since become a foundational architecture in modern AI and ignited the current AI boom by enabling the era of large language models (LLMs).

Leveraging the Transformer architecture and massive training corpora, researchers pretrained extremely large networks (with billions of parameters) on language modeling objectives. Notable examples include "BERT" [90] and the GPT series [18], [91], which are pre-trained on vast text data and then fine-tuned for specific tasks. These LLMs achieved unprecedented performance across a wide range of language understanding and generation tasks, often matching or exceeding human-level benchmarks on question answering, summarization, and more. For instance, BERT achieved the best results ever recorded (at that time) on 11 widely used tests that measure language understanding, reflecting a new versatility in AI systems. Therefore, it is not surprising that LLMs are now being explored for their ability to aggregate and reason over knowledge. Modern LLM-based systems can ingest documents or web content and answer factual questions, essentially serving as automated research and fact-checking agents. This capability suggests that large language models, when properly constrained and verified, could operate as AI Oracles that provide reliable natural-language answers to blockchain smart contracts. Indeed, recent work has demonstrated prototype "AI oracles" that use LLMs to automatically source and verify information from diverse online sources [92]. Such systems combine advancements in NLP with decentralized consensus mechanisms to aim for trustworthy data feeds.

**2.8 Adversarial Machine Learning and Generative Models**

As AI systems became more capable, researchers also uncovered vulnerabilities and new challenges. One major development is the field of adversarial machine learning, which studies how malicious inputs or perturbations can fool AI models and how to make models more robust. Szegedy et al. [93] first revealed that even imperceptibly small changes to an input (e.g. an image) could cause a confident neural network to misclassify. This sparked a wave of research into adversarial attacks and defenses. Goodfellow et al. [20] introduced the fast gradient sign method (FGSM) to generate such adversarial examples efficiently, and subsequent work showed attacks were possible even without internal model access (black-box attacks) and could be made robust to real-world conditions. In response, numerous defense strategies have been proposed (adversarial training, input sanitization, verification techniques), but achieving fully robust models is still an open challenge. The adversarial ML literature is highly relevant to blockchain oracles because an oracle mechanism might be targeted by adversaries providing specially crafted data designed to mislead an AI-based oracle. Ensuring adversarial resistance, that is, the ability to detect or withstand maliciously manipulated inputs, is crucial if AI models are to be trusted in a decentralized oracle context.

Anomaly detection is for example a practical method to ensure reliability flagging or rejecting anomalous data. Anomaly detection is a well-established subfield of ML that focuses on identifying outliers or unusual patterns that do not conform to expected behavior [94]. This has critical applications in fraud detection, network intrusion detection, fault monitoring, and more, domains analogous to the oracle setting where an anomalous data point might indicate faulty or malicious input. A variety of techniques exist (statistical tests, clustering-based methods, one-class SVMs, autoencoder networks, etc.), but generally they model what "normal" data looks like and then measure deviations

Another innovation of this era is the advent of Generative Adversarial Networks (GANs) [95]. GANs consist of two neural networks a generator and a discriminator locked in a competitive game. The generator tries to create synthetic data (images, for example) that are so realistic the discriminator cannot tell them apart from true data, while the discriminator improves at spotting fakes. This adversarial training process enables GANs to produce remarkably realistic outputs, effectively learning the true data distribution. GANs revolutionized generative modeling and have been used in contexts from image synthesis to data augmentation. In the oracle problem space, GANs might not be directly used to generate oracle data, but their emergence underscores how AI can now create extremely realistic fake data raising the stakes for truth verification. For instance, GAN-generated "deepfake" content (images, text) could fool naive oracles, hence oracle designs must account for the possibility of highly realistic but false data inputs. On a positive note, adversarial training concepts could also be leveraged to design an oracle AI that actively anticipates deceptive inputs and is trained to be skeptical, much like a GAN's discriminator is trained to spot fakes.

These foundational developments in artificial intelligence, including symbolic reasoning, machine learning, reinforcement learning, natural language processing, and adversarial robustness provide the conceptual and technical tools relevant to the design of AI-assisted oracle systems. In the following sections, we examine how these AI capabilities intersect with the blockchain oracle problem, drawing on both current research initiatives and practitioner proposals, as well as informed speculation on possible future integrations. The following table (Table 2) summarizes the information provided above.

Table 2. Overview of AI Paradigms

| Year Range | AI Paradigm | Main Advancement | Drawbacks | Reference |
|---|---|---|---|---|
| 1950s-1970s | Symbolic AI / Expert Systems | Rule-based reasoning and knowledge encoding | Lacks generality; brittle; hard to scale rule sets | [65], [67] |
| 1980s-1990s | Connectionism / Neural Networks | Multi-layer learning via backpropagation | Unstable training; limited depth | [96], [97] |
| 1990s-2000s | Statistical Machine Learning | Probabilistic models; optimization-based learning | Shallow features; limited in unstructured data tasks | [71], [98] |
| 2000s-2010s | Deep Learning | End-to-end learning from large datasets | Opaque models; data-hungry; high compute cost | [99], [100] |
| 2010s-Present | Reinforcement Learning | Learning via rewards and interactions | High complexity; reward shaping required; sample inefficiency | [101], [102] |
| 2010s-Present | NLP & Large Language Models | Transformer models for understanding text | Prone to hallucinations; computationally intensive; non-deterministic | [18], [89], [103] |

*Author elaboration

## 3. What Can AI Do to Support Oracle Systems

As blockchain applications continue to grow in complexity and criticality, ensuring the reliability, accuracy, and responsiveness of oracles becomes increasingly vital. Artificial Intelligence (AI) offers a broad spectrum of techniques that can enhance oracle systems across multiple dimensions, from anomaly and adversarial behavior detection to intelligent node selection, automated fact extraction, and the integration of hybrid AI-governance models. This section examines the various roles AI can play in enhancing oracle functionality, analyzing recent academic and industry developments that aim to strengthen oracles against manipulation, inefficiency, and unreliability.

### 3.1 AI for Anomaly Detection in Blockchain Oracles

As thoroughly explained in the introduction, incorrect or manipulated oracle data can lead to severe consequences, including financial losses and compromised smart contract executions. Risks can stem from both benign anomalies, such as sensor errors or network delays, and intentional adversarial behaviors, like flash loan-induced price manipulations. While anomalies typically result from unintended technical failures, adversarial manipulations are deliberate actions by malicious actors exploiting oracle vulnerabilities. AI and Machine Learning (ML) have emerged as pivotal tools in detecting, analyzing, and mitigating these heterogeneous risks to enhance oracle security and reliability.

Statistical anomaly detection, for example, utilizes AI to identify data points or behaviors significantly deviating from expected norms, primarily due to non-malicious technical errors or unexpected external events. Techniques include simple statistical filtering (median or mean-based outlier rejection), clustering algorithms, isolation forests, and autoencoders. For instance, decentralized oracle networks like Chainlink apply basic statistical methods to aggregate data from multiple nodes, identifying outliers when submissions substantially diverge from the median consensus. If most nodes report similar values and only a few differ significantly, these inconsistent submissions are discarded or marked for additional verification [104].

Advanced statistical methods, such as Long Short-Term Memory (LSTM) autoencoders, enhance anomaly detection by capturing complex temporal dependencies in oracle data streams. By modeling historical price feeds, these deep learning models predict expected values and flag significant deviations as anomalies [105]. Using data from Band Protocol, for example, researchers have shown that LSTM autoencoders successfully detected abnormal price fluctuations, providing robust alerts against unusual but non-adversarial market movements. Similarly, Park et al. [106] utilize hybrid statistical methods such as Kalman filters combined with conformal prediction to update uncertainty intervals dynamically. When real-time oracle inputs deviate beyond these intervals, they are flagged as potential anomalies, requiring manual or additional automated verification. Such AI-driven statistical techniques significantly strengthen oracle systems against unpredictable data inconsistencies.

In this context, unsupervised or semi-supervised anomaly detection is very powerful as it assumes that "most data is normal" and flags anything sufficiently deviant. An advantage of unsupervised methods is that they can catch previously unseen anomalies, although they require careful tuning to avoid false positives in highly variable data like crypto prices. Combining data from multiple assets or sources can be a strategy to enhance these systems. For example, a model might consider not just one price feed but also related market indicators (volume, broader market movement) to judge if a price change is anomalous in context. Advanced models might use graph neural networks or correlation analysis across multiple feeds as an oracle often outputs many data points (for different trading pairs, etc.). Anomalies might be more evident when considering the whole graph of assets (e.g., if only one asset out of many moves 50% while others move 1%, that could be flagged) [107]. For instance, Ikeda *et al*. [108] propose an anomaly indicator that fuses many metrics (entropy, clustering coefficients, etc.) using a Boltzmann machine, though in the context of crypto trading anomalies. Translating such multi-metric approaches to oracle data could mean examining not only values but also node network metrics together.

**3.2 Detection of Adversarial and Manipulative Behavior**

While statistical anomaly detection addresses primarily benign errors, adversarial detection specifically targets intentional malicious data manipulations including flash loan attacks or Sybil attacks. These attacks exploit vulnerabilities to deliberately distort oracle inputs, causing significant financial damage to DeFi platforms.

Abinivesh et al. [107] demonstrated that AI-driven oracles offer measurable gains in addressing adversarial behavior. For instance, one prototype that combined multi-source aggregation with an RL-based decision agent achieved 92% fraud detection accuracy, substantially higher than a traditional non-learning oracle's 78% accuracy. The RL enhanced oracle dynamically adjusted trust scores and could "learn" to reject bad data, resulting in a false-positive rate (rejecting good data) of only 4%, versus 12% in a legacy oracle network. Moreover, adding an AI-powered fraud detection module (e.g., an anomaly classifier watching for unusual submission patterns) can boost accuracy even further, as one hybrid model reported 94% accuracy with only 2% false positives.

Recent frameworks, such as AiRacleX, further utilize advanced large language models to automatically detect price oracle manipulation attempts in decentralized finance protocols. AiRacleX operates by first extracting comprehensive knowledge about known vulnerabilities and attack patterns from blockchain security literature, then employing targeted prompting techniques to analyze smart contract logic proactively. Through extensive empirical testing against real-world exploits, AiRacleX significantly outperformed traditional detection methods, providing enhanced recall rates and precise identification of malicious behaviors [15].

More complex manipulation, such as Flash loan attacks, instead represent a critical threat, as they leverage instantaneous, high-volume borrowing to artificially inflate or deflate asset prices temporarily [109], [110]. The Mango Markets exploit of 2022 exemplifies such threats, where attackers manipulated oracle price feeds to borrow excessive funds against artificially inflated collateral [111]. AI-driven detection models, particularly supervised learning methods, have proven effective in identifying and mitigating these attacks by analyzing intricate transaction patterns in real time. For example, detection systems such as Forta implement heuristic or ML-based detectors to recognize typical flash-loan attack sequences and trigger protective measures like halting a protocol or rejecting an oracle update [112].

Sybil attacks, on the other hand, involve adversaries controlling multiple oracle nodes to artificially influence consensus outcomes. A group of nodes (possibly Sybils controlled by one entity) can, in fact, feed the same wrong data, making an outlier check difficult. AI can aid in detecting correlated anomalies that suggest collusion. For instance, since nodes disagree occasionally due to random error, but suddenly a subset of nodes all move in unison to a new value that others do not, that pattern, as explained in [107] might be caught by a clustering or graph-based anomaly detector [113]. Abinivesh et al. also supports the possibility of preventing sybil behavior by analyzing oracle timing or semantics. As oracles are meant to operate independently, if a subset always submits their data within the same millisecond or with identical metadata, this could indicate a single operator behind them.

### 3.3 AI for Oracle Node selection.

As seen in the previous paragraphs, ML and statistical analysis are utilized to discard outliers and ensure that data is coherent, favoring historically reliable data sources. Recent research shows however that this data is not leveraged by oracle providers in real-time who generally use instead static data, creating a potential bias in node selection [107]. AI can enhance these mechanisms by dynamically scoring data quality instead of relying on static thresholds. Taghavi et al. [114], for example, employs Bayesian reinforcement learning frameworks to dynamically adjust oracle node reputations, leveraging real-time performance metrics such as accuracy, responsiveness, and reliability. Nodes demonstrating consistent reliability receive higher reputation scores, incentivizing honest reporting. Conversely, nodes exhibiting erratic or suspicious behaviors receive lower scores, effectively isolating potentially compromised nodes. Experimental implementations on Ethereum demonstrated BLOR's effectiveness in consistently identifying optimal oracle nodes, significantly reducing operational risks and costs.

Similarly, Zhang *et al*. [115] introduced a deep reinforcement learning model (TCODRL) that incorporates a comprehensive trust management framework. It evaluates oracle reputation on multiple dimensions using a sliding window to track changes, and then applies deep RL to adaptively select high-reputation oracles. In simulations, this system reduced the usage of malicious oracles by >39% and cut overall costs up to 12% compared to traditional static methods. These results suggest AI can significantly improve oracle data by learning which data sources tend to be honest or accurate.

Other proposals for reputation schemes consider multi-dimentional signals. For instance, the ETORM proposal tracks each oracle's task-level accuracy and completion time (local reputation) and its overall historical performance and uptime (global reputation). Oracles commit stake that can be slashed on misreports [116]. These metrics are combined (often weighted by recency via a sliding window) into a single trust score used to filter and rank nodes [115]. In principle, ML could further refine this by learning which features best predict reliability. For example, clustering or outlier-detection could spot anomalous oracle behavior.

**3.4 Hybrid AI-Governance Models for Oracle Reliability**

While AI techniques provide robust and dynamic methods for evaluating oracle reliability, these techniques are most effective when integrated with decentralized governance frameworks and cryptoeconomic incentives. Rewarding and punishing agents for complying with specific operations is also a principle of Reinforcement learning that can be efficiently implemented in governance mechanisms in light of balancing algorithmic accuracy and community-driven decision-making.

An example of this integration is Supra's Threshold AI framework that requires each AI agent (an oracle node running an AI model) to lock a stake and earn a performance-based reputation. If an agent produces incorrect or malicious outputs, the protocol will slash its staked collateral as a penalty. Otherwise, agents that consistently provide timely, accurate data are rewarded with user fees or token subsidies. By embedding staking, slashing, and reputation scores at the core of the oracle, the system creates financial disincentives for bad data and drives AI agents to act honestly. Such cryptoeconomic guarantees are crucial, given that AI models could otherwise behave in an opaque manner. The stake, on the other hand, provides a tangible accountability for the AI operator. Notably, a sufficiently large stake also raises the cost of Sybil attacks (spawning fake oracle nodes) and can even serve as a trust signal (long-duration or high-value takes increase an agent's reputation weight). This incentive based system is quite known and widespreadly used in the world of oracles since early days of Ethereum by operators such as Tellor, Razor or Bluzelle [49], [117], [118]. However ensuring proactiveness and responsiveness of human nodes to economic incentives is not always measurable due to laziness and limited action flow [24], [58]. Thanks to RL, AI agents can be efficiently trained with economic incentives, and their behavior can be fairly predictable.

Decentralized Autonomous Organizations (DAOs) may also play a crucial role in complementing AI-driven reliability models. Oracle providers such as API3, for example, utilize decentralized governance to allow stakeholders to vote on critical oracle management decisions, including adding or removing data sources, adjusting update frequency thresholds, and managing network parameters [45]. DAO's decisions and updates may help rebalance AI parameters so that the implemented models or agents are adapted to the protocol's needs and market changes. Ironically, human oversight remains an important backstop in some AI oracle proposals. Because AI agents might struggle with subjective or ambiguous queries, a "human-in-the-loop" mechanism can be used as a last resort. The Supra framework, for example, allows certain queries to be flagged for manual review. If the AI committee can't reach a confident consensus, the query can escalate to designated human arbiters or a DAO vote before finalizing the on-chain result [12]. Humans can either override the AI's output or participate alongside AI agents in consensus for those cases. While this introduces some latency, it provides a crucial check on AI decisions and helps handle things that algorithms can't or shouldn't decide alone. This is again a well-known and established principle in the oracle space for solutions such as RealityEth, Augur or UMA, where for complex or delicate decision, the protocol escalates to an external arbiter (Kleros), which is notably human-based [43], [48], [119]. In effect, decentralized human consensus acts as the ultimate oracle. Finally, we can argue that the combination of AI automation with community governance and staking creates a hybrid trust model in which AI brings speed and scalability in analyzing data, while decentralized human and economic mechanisms provide accountability, configurability, and fallback in cases where AI might err or be uncertain.

### 3.5 AI-Driven Fact Extraction and Verification in Oracle Systems

Natural Language Processing (NLP) techniques and Large Language Models (LLMs) are increasingly proposed as tools to assess the trustworthiness of unstructured or semi-structured data before it is submitted on-chain. Traditional oracles often relay raw data (prices, event outcomes, etc.) without interpretation, but LLM-powered oracles could interpret and verify facts from sources like news articles, financial filings, or weather reports. For example, Chainlink Labs investigated an oracle prototype that uses an LLM to parse corporate reports and press releases for specific events (e.g. dividend announcements) and convert them into a structured format [13]. In their tests, multiple oracle nodes ran independent LLM instances to cross-verify the extracted facts, helping filter out hallucinations and errors. Only when the nodes reach consensus on a fact (e.g. the exact dividend amount and date) is the information accepted and published on-chain. Likewise, evidence from practitioner research supports the view that LLM agents can autonomously retrieve documents, analyze content, and even cite sources as evidence for claims [22]. By grounding their outputs in verifiable references and providing reasoning traces, such systems aim to ensure each on-chain fact is backed by transparent evidence, increasing confidence in the oracle's data.

Beyond data retrieval, LLMs can act as an inference layer within decentralized oracles, performing reasoning or judgment on incoming data. Rather than simply reporting an external value, an LLM-enhanced oracle could answer complex queries like *"Did a certain regulatory change actually occur?"* or *"Should a liquidation execute given current market news?"*, returning a yes/no or contextual answer that has been vetted by AI reasoning [120]. A recent implementation in the practitioner space proposes to achieve this through multiple LLM-based agents with different roles to deliberate over an event and reach a quorum before delivering an outcome. Agents may also embody different roles from simple fact checking to data inconsistencies or legal compliance, while their collective decision may be aggregated once a threshold is reached and a cryptographic proof is generated (i.e., BLS signature) for the result [121].

Concerning complex queries, again Chainlink research team built an LLM-based prediction market resolver that autonomously determined real-world event outcomes for Polymarket markets. Using GPT-4 with a carefully designed pipeline (question reframing, web research via tools like Perplexity, and a reasoning module), their AI oracle correctly resolved ~89% of 1,660 test cases, even citing sources for each answer. Intuitively, it excelled in cases with clear official data (sports results, etc.) and logged a transparent chain-of-thought for auditing [22].

Pioneering work has also been pursued by Oraichain, which launched a specialized blockchain that is meant to act as an AI-centric oracle. It allows smart contracts to access various AI models including LLMs, for data analysis, content moderation, and verification. They also developed an interesting method to enhance and verify the reliability of AI responses, leveraging test cases and having AI vote on these cases. To make an example, before AI oracles are used to answer on a real use cases, a test query is run and the answers are verified through a benchmark that identifies which agents are reliable and can intervene in the real use case [14]. An example of this system is "Modestus", a content moderation oracle built on Oraichain that uses an LLM to classify text under various policies (hate speech, profanity, etc.). Modestus was trained by drawing knowledge from multiple black-box LLMs into one open-source model, using a decentralized aggregation of their outputs to reduce individual model bias [122]. This allows for the reduction of the blurriness of LLM model reasoning while permitting adjustments if deemed necessary. The higher level of transparency may also allow a more agile auditability.

Academic research reinforces the idea of leveraging LLM models to improve oracle reliability. For instance, Xian *et al.* [16] introduce C-LLM, a framework where multiple oracle nodes query independent LLMs and then apply a truth-discovery algorithm (called SenteTruth) to aggregate the answers. By combining semantic similarity measures with voting/truth-detection methods, they showed improved answer accuracy up to 17.7% even with nearly 40% of nodes being malicious or unreliable. Xian et al. approach treats LLMs as a decentralized validator, counting as a single voter instead of a middle layer solution. That way, allucination-driven errors are highly mitigated [16]. This study directly extends a pioneering research by Xu et al. [123], which proposed a system for smart contracts to query LLMs using a relayer. In their system, smart contracts and LLM worked independently, and a verification mechanism ensured the relayer couldn't tamper with the LLM response (either with a hash comparison or with a ZKP). They also proposed a wrapper at the smart contract level that allows for formatting questions and interpreting LLM responses effectively. This work is particularly important as being also blockchain agnostic, may serve as a trailblazer for further research, such as Xian et al. [16]. Table 3 provides a summary of the information discussed in this section.

Table 3. AI Paradigms, Techniques, and Use Cases in Oracle Systems

| Topic | Description | Techniques | Applications and References |
|---|---|---|---|
| AI for Anomaly Detection in Oracle Data | Detection of unexpected deviations in oracle inputs (due to technical errors or market fluctuations) using AI to enhance data reliability and prevent smart contract malfunctions. | Statistical filtering, isolation forests, LSTM autoencoders, Kalman filters, conformal prediction, unsupervised learning, graph-based models, and Boltzmann machines. | Chainlink (median filtering), [52] Band Protocol (LSTM), [46] Park et al. [106], Kalman + conformal, (Boltzmann fusion), multi-asset correlation models. [108] |
| Detection of Adversarial and Manipulative Behavior | AI methods are used to detect deliberate attempts to manipulate oracle data (e.g., flash loans, Sybil attacks), enhancing oracle resilience against targeted exploits. | Reinforcement learning, supervised learning, clustering, graph-based detection, LLM-based reasoning, and temporal and semantic correlation analysis. | RL oracle [107], AiRacleX (LLM detection), [15] Forta (flash loan patterns), [112] Sybil detection via clustering [113]. |
| AI for Oracle Node Selection | AI enhances dynamic selection of oracle nodes by scoring them in real-time based on reputation, accuracy, and reliability, reducing dependency on static configurations and mitigating selection bias. | Bayesian reinforcement learning, deep reinforcement learning, trust scoring, sliding window analysis, clustering. | BLOR [114], TCODRL [115], ETORM [116] |
| Hybrid AI-Governance Models for Oracle Reliability | Combining AI-driven evaluation with decentralized governance and cryptoeconomic incentives improves oracle reliability by aligning automated decision-making with community oversight and financial accountability. | Reinforcement learning, staking and slashing, reputation systems, human-in-the-loop escalation, DAO-based governance. | Supra's Threshold AI [12], API3 DAO governance [45], Augur [43], Kleros arbitration [119], Tellor [49] |
| AI-Driven Fact Extraction and Verification | LLMs and NLP models are used to autonomously retrieve, interpret, and verify facts from unstructured sources (e.g., news, filings) before committing data on-chain. Cross-verification, grounding in source documents, and transparency mechanisms aim to enhance trust and accuracy. | Large Language Models (LLMs), semantic similarity voting, reasoning traces, multi-agent deliberation, cryptographic proof aggregation (e.g., BLS signatures), benchmarking, and role-based agent scoring. | Chainlink LLM oracle prototype [13], Supra's prediction market resolver [120] Oraichain's "Modestus" content moderation oracle [122], C-LLM and SenteTruth [16], LLM-query relayer framework [123] |

## 4. Challenges for AI in solving the oracle problem

While AI technologies offer promising enhancements to oracle systems, their integration into decentralized blockchain infrastructures remains fraught with critical challenges. The present section provides a structured and critical overview of these limitations, highlighting technical, epistemological, and governance-related obstacles. Particular emphasis is placed on how AI's inherent characteristics, such as non-determinism, opacity, and data dependency, may conflict with blockchain principles like verifiability, trust minimization, and deterministic consensus.

## 4.1 Lack of Cryptographic Verifiability and Determinism

Blockchain technology's foundational strength lies in its deterministic and cryptographically verifiable nature. identical inputs must consistently yield identical outcomes across all nodes, ensuring universal consensus and trustless verifiability [124], [125]. Sophisticated AI models instead, particularly deep neural networks and large language models (LLMs), possess probabilistic and non-deterministic behaviors, making integration in purely decentralized oracle architectures inherently complex. AI models commonly incorporate randomized elements such as stochastic gradient descent, model initialization, and sampling procedures, leading to probabilistic outputs [18], [89]. Consequently, even identical AI setups across different blockchain nodes may produce slightly different results, undermining the consistency and unanimity required by blockchain consensus mechanisms [126], [127].

The non-determinism inherent in LLMs further complicates these consensus challenges. For instance, large language models may produce variable outputs on repeated queries due to their generative and probabilistic nature, especially when parameters like sampling temperature are not strictly controlled [128]. Temperature is a parameter that controls how random or deterministic the sampling process is. Fixing this parameter to zero can indeed reduce randomness; however, such constraints can negatively affect model flexibility and output quality, indicating an inherent trade-off between determinism and model performance [17]. Decentralized oracle architectures employing LLMs thus require additional and sophisticated mechanisms that ensure univocal responses (e.g., SenteTruth[16]), introducing additional complexity, but arguably without fully eliminating ambiguity. In cases where high-confidence consensus is unreachable, oracles might need to acknowledge query indeterminacy explicitly, further complicating integration [22].

The opacity of AI decision-making also clashes with blockchain's transparency and auditability principles [129], [130]. Complex AI systems, especially deep neural networks, often function as "black boxes," lacking fully transparent reasoning pathways [131], [132]. This opacity creates substantial trust and governance issues, particularly critical in high-stakes blockchain applications such as finance, governance, or legal agreements, where verifiability and explainability are crucial [133], [134], [135]. While some AI oracle design proposals incorporate explicit reasoning logs or cryptographic quorum poofs, the verification of AI-generated outputs requires inspecting these transcripts off-chain, weakening or impeding full on-chain auditability and introducing additional off-chain trust reliance mechanisms [12], [22].

In practical terms, the discrepancy between blockchain determinism and AI probabilism necessitates supplementary verification measures. Current strategies exploring verifiable computation for AI, including zero-knowledge proofs (zk-SNARKs or zk-STARKs), are still nascent, computationally expensive, and largely impractical for large-scale models or real-time applications [16], [123]. While emerging oracle systems (e.g., Oraichain) attempt transparency through open-sourced models and verifiable inference processes, the complexity and resource-intensiveness of such solutions presently limit their widespread adoption and scalability [14]. Table 4 summarizes what was discussed in the present paragraph.

Table 4. Lack of Cryptographic Verifiability and Determinism: Core Challenges and Implications for Oracles

| Core Challenges | Underlying causes | Implications for oracles | Key References |
|---|---|---|---|
| AI models (especially LLMs) produce non-deterministic, probabilistic outputs. | Use of stochastic training methods (e.g., random initialization, sampling temperature). | Non-deterministic outputs can disrupt node consensus.<br><br>Full on-chain verification of AI outputs is infeasible. | [14], [16], [18], [89], [124], [128] |
| Blockchain requires deterministic execution for consensus and verifiability. | Models behave like "black boxes," with limited transparency or explainability. | Additional mechanisms (e.g., SenteTruth, zero-knowledge proofs) add complexity and cost.<br><br>Off-chain trust or fallback systems may reintroduce centralization. | |
| This mismatch creates fundamental integration problems. | Difficult to replicate outputs exactly across nodes. | Practical limitations in real-time and large-scale deployment due to computational intensity. | |

### 4.2 Model Fallibility and Bias

Despite significant advancements, AI models remain inherently fallible, susceptible to biases, and prone to systematic inaccuracies, presenting evident limitations within blockchain oracle applications. Particularly critical are issues of false positives and false negatives arising from anomaly detection systems. For instance, legitimate market movements characterized by exponential but authentic price fluctuations can be misidentified as anomalous events (false positives), potentially resulting in unnecessary disruptions or delays to smart contract processes. Conversely, carefully crafted adversarial inputs may exploit known weaknesses in AI models, resulting in overlooked malicious manipulations (false negatives) [136], [137]. This challenge becomes critical in highly volatile environments like decentralized finance (DeFi), where sensitive AI systems must delicately balance alert thresholds to minimize both types of errors [114], [115]. For instance, we would like to clarify that we are not claiming that AI oracles have already demonstrably failed due to false positives or false negatives, but given the well-documented limitations of AI-based anomaly detection in high-volatility and adversarial environments [20], it is reasonable to infer that AI-enhanced oracle systems remain vulnerable to false positives and false negatives, particularly in fast-moving DeFi markets.

Further complicating these issues, large language models (LLMs), despite their powerful reasoning capabilities, are particularly prone to "hallucinations", outputs that appear plausible but contain entirely fabricated or unsupported information [18]. Within blockchain contexts, reliance on hallucinated information can trigger erroneous automatic executions in smart contracts, potentially causing financial losses, improper settlements, or legal disputes. To mitigate hallucinations, approaches like robust source-grounding, explicit reasoning traces, and cross-verification through multiple models have been explored [14],

[22]. However, as LLMs mostly share the same weaknesses and are potentially trained on the same datasets, they would probably all converge to a hallucination if they are prompted to produce an output on a segment of data that is lacking.

Additionally, AI models inherently embed biases present in their training data. If an oracle's AI model has been predominantly trained on historical data from a specific market, region, or provider, it may systematically underperform or inaccurately assess data originating from novel or underrepresented contexts [19], [138]. To make an example, if an anomaly detection model is trained on a specific DeFi market, when implemented in another market, it may potentially misreport anomalies or overlook manipulations. Therefore, it may require additional training data and testing before performing well in another context.

AI models also face challenges related to model drift and degradation over time, as real-world data distributions evolve and adversaries exploit newly discovered vulnerabilities [139], [140]. For example, a predictive AI oracle initially performing well may gradually lose accuracy if its training data no longer represent current market conditions or if adversaries engineer inputs to deceive it [12]. Continuous retraining and dynamic model updates become necessary to sustain accuracy; however, implementing these updates in decentralized systems may require complex governance processes or multi-party verification, which adds layers of procedural complexity and potentially delays critical updates. Otherwise, the AI implementation must be managed by a centralized entity, which clashes with blockchain decentralization principles.

At the end of the day, despite artificial intelligence's considerable strengths, its reliability is strictly dependent on the trustworthiness of external data sources, a challenge well-known in computer science as the "garbage-in, garbage-out" (GIGO) principle. This principle emphasizes that the output quality of any computational system, regardless of sophistication, directly depends on the accuracy and authenticity of the input data it receives [141], [142]. No matter how advanced or intricate an AI model may be, it remains constrained by the veracity and integrity of the initial data provided.

This limitation becomes particularly problematic in decentralized blockchain systems, where the main goal is trustlessness and independent verifiability [5], [143]. AI-driven oracle solutions, as any other oracle, although significantly enhancing data reliability "under ideal conditions", inherently rely on external information sources. These data sources such as sensor networks, financial market feeds, or third-party reports, are beyond the blockchain's native verification capabilities [4], [6]. For example, while AI-based anomaly detection methods can effectively flag suspicious price fluctuations or irregular data submissions, they cannot independently verify the accuracy of these data points [136], [144]. Similarly, natural language processing (NLP)-based oracles, which extract structured facts from textual content, depend entirely on the trustworthiness and accuracy of their primary information sources. Consequently, if the original documents or sources contain inaccuracies or misinformation, the AI systems will inadvertently propagate and amplify these errors, producing sophisticated but ultimately flawed conclusions [17], [145], [146].

Therefore, While AI can reduce risks associated with data inaccuracies, it cannot eliminate the need for external trust in information sources. Table 5 summarizes these concepts

Table 5. Model Fallibility and Bias: Core Challenges and Implications for Oracles

| Core Challenges | Underlying causes | Implications for oracles | Key References |
|---|---|---|---|
| AI models are inherently imperfect: they can hallucinate, misclassify, or underperform in new contexts.<br><br>They are vulnerable to false positives/negatives, data drift, hallucinations, and biases from training data. | High volatility in DeFi markets increases the difficulty of precise anomaly detection.<br><br>LLMs may generate plausible but incorrect content ("hallucinations").<br><br>Models trained on narrow datasets may fail in new domains (data bias).<br><br>Over time, changing data patterns lead to model drift or degradation.<br><br>GIGO principle: AI output is only as good as the data input. | False alerts or missed threats may lead to smart contract failures.<br><br>Hallucinated facts could trigger erroneous on-chain actions.<br><br>Biased or outdated models may misreport data in unfamiliar conditions.<br><br>Decentralized retraining and updating are difficult and slow.<br><br>Full trustlessness cannot be guaranteed, as data source trust is still needed. | [12], [17], [136], [137], [140], [142], [144] |

## 4.3 Complexity and Expanded Attack Surface

Integrating advanced AI techniques into blockchain oracle systems significantly increases both architectural complexity and the potential attack surface, introducing new vulnerabilities alongside enhanced capabilities. As emphasized in previous paragraphs, sophisticated AI models, including neural networks, large ensembles, and reinforcement learning frameworks, inherently demand considerable computational resources for training, fine-tuning, and inference [18], [144], [147], [148]. Due to blockchain's stringent on-chain resource constraints (such as gas costs and computational limitations), such advanced computations often require off-chain execution coupled with secure transmission and cryptographic verifications back to the blockchain [12], [22]. This additional operational layer introduces complexities around data transmission protocols, verification methods (e.g., zero-knowledge proofs or trusted execution environments), and ensuring the integrity and authenticity of off-chain computation results, significantly complicating the architecture and potentially introducing latency and scalability bottlenecks [16], [123]. In practical terms, it means that the information needed and delivered by an AI oracle should be transmitted to the blockchain through another oracle, which is a very controversial solution in light of decentralization and intermediary reduction. Moreover, AI integration inherently exposes oracle systems to adversarial machine learning attacks. Techniques such as data poisoning, intentionally corrupting training datasets to produce systematically flawed outputs, and adversarial input manipulations designed to deceive AI models represent tangible, severe threats [21], [149]. For instance, carefully constructed adversarial inputs or prompts can exploit the sensitivity of large language models (LLMs), eliciting biased, incorrect, or misleading outputs, potentially triggering harmful or erroneous blockchain actions. It has to be considered that robustness testing through simulated adversarial scenarios, adversarial training, and the employment of ensemble models to cross-validate outputs becomes essential, thus inevitably affecting implementation costs [20], [150].

Additionally, the complexity of AI-driven oracle architectures significantly complicates security auditing, validation, and operational monitoring of smart contracts. Every new AI component introduced into the oracle system demands rigorous security assessments and continuous verification, each carrying substantial costs in terms of expertise, resources, and time. Subtle coding errors, overlooked edge cases, or unforeseen model behaviors could lead to severe vulnerabilities that malicious actors can exploit, significantly expanding the potential attack vectors beyond those of simpler, traditional oracles. Bugs in smart contracts, flawed oracle architectures, and unforeseen events have already led to dramatic failures in blockchain history. For instance, the DAO hack that resulted in the Ethereum hard fork was caused by a smart contract vulnerability (re-entrancy); the Curve Finance incident stemmed from poor oracle selection; and the mass liquidations in 2020 were triggered by a black swan event (COVID-19), for which no adequate safeguards had been implemented [58], [151], [152]. Introducing off-chain AI components into oracle systems would likely increase architectural complexity and, consequently, expand the potential attack surface, potentially leading to more frequent or severe failures of the kinds described above. While it is true that these historical failures were rooted in human error, the same applies to AI: before it is artificial or autonomous, it is programmed, trained, and managed by humans [153], [154].

Consequently, the integration of advanced AI techniques into blockchain oracle systems, while offering notable advantages, demands cautious, strategic implementation. Rigorous calibration, comprehensive security audits, continual adversarial testing, and careful balancing of complexity against performance remain indispensable. Ultimately, the practical deployment of AI-enhanced oracles must critically evaluate whether their security benefits genuinely outweigh the substantial operational and security overheads they introduce, particularly within decentralized contexts that prioritize transparency, trustlessness, and broad accessibility. The following table (Table 6) provides an overview of the information discussed in this paragraph.

Table 6. Complexity and Expanded Attack Surface: Core Challenges and Implications for Oracles

| Core Challenges | Underlying causes | Implications for oracles | Key References |
|---|---|---|---|
| Integrating advanced AI introduces architectural complexity and widens the attack surface.<br><br>New layers (off-chain computation, model training, verification protocols) create vulnerabilities and operational friction.<br><br>AI systems are susceptible to adversarial machine learning threats and are harder to audit. | AI models require high computational resources, often incompatible with blockchain's on-chain constraints.<br><br>Off-chain execution introduces reliance on additional oracles and complex verification schemes (e.g., ZKPs, TEEs).<br><br>Adversarial attacks (e.g., data poisoning, crafted inputs) can exploit AI models.<br><br>Increased system complexity makes auditing and monitoring more difficult, amplifying the risks of bugs or misbehavior. Failures from poor oracle design or smart contract bugs in the past (DAO hack, Curve incident, 2020 liquidations) highlight the risks of complexity. | Off-chain AI requires secure relay mechanisms, possibly undermining decentralization goals.<br><br>Robustness testing becomes essential but costly.<br><br>System complexity can slow deployment, increase maintenance burdens, and create more failure points.<br><br>Human error remains a root cause. It's not eliminated by AI, it's just shifted. | [18], [22], [123], [149], [150], [151], [153] |

## 5. Conclusive thoughts

*"If we use, to achieve our purposes, a mechanical agency with whose operation we cannot efficiently interfere…, we had better be quite sure that the purpose put into the machine is the purpose which we really desire."* (Wiener, 1960 [155])

These words from Norbert Wiener remain strikingly relevant in today's discussions around both artificial intelligence and blockchain oracles. When we delegate decision-making to external systems, whether deterministic or probabilistic, we must ensure that their internal logic aligns with our objectives and values. Failing to do so risks not only inefficiency but profound systemic failure.

The blockchain oracle problem is not just a technical limitation; it is an epistemic one. It reflects the paradox of attempting to create trustless systems that ultimately depend on data whose authenticity cannot be independently verified. In that sense, the problem is not eliminated but merely displaced: from verifying data to verifying data providers, from trusting central parties to trusting probabilistic mechanisms or game-theoretic assumptions.

This paper has explored the potential role of artificial intelligence in addressing this dilemma. As shown through the analysis of current research and implementations, AI can meaningfully support oracle infrastructures, enhancing anomaly detection, ranking data sources dynamically, interpreting unstructured information through NLP, and detecting manipulation with adversarial learning techniques. Frameworks such as AiRacleX and industry protocols like Chainlink and Oraichain demonstrate how AI can be embedded into oracle systems to expand their analytical capabilities [14], [15], [22].

Yet, this integration does not resolve the oracle problem. AI does not remove the need for trust; it redistributes it. It introduces new forms of opaqueness, shifts the point of failure, and adds layers of complexity that must themselves be monitored, audited, and secured. In essence, trying to solve one black box with another is conceptually incoherent. While AI can optimize, it cannot verify truth in a cryptographically meaningful way.

Therefore, the most reasonable path forward lies in hybrid architectures: oracle systems that strategically combine AI-powered inference with economic incentives, decentralized governance, cryptographic proofs, and transparent accountability mechanisms. These systems should not aim to eliminate trust altogether but to manage and distribute it in ways that are auditable, resilient, and context-appropriate.

The title of this paper poses a provocative question: *Can AI solve the blockchain oracle problem?* After careful consideration, the answer is clearly no, but it can help mitigate it. Recognizing the limits of both technologies allows us to design oracle systems that are neither naively deterministic nor recklessly optimistic. As the space continues to evolve, what is needed is not technological absolutism, but pragmatic innovation grounded in interdisciplinary rigor and epistemic humility.